\documentstyle[12pt]{article}
\newcommand{\be}{\begin{equation}}
\newcommand{\ee}{\end{equation}}
\newcommand{\bea}{\begin{eqnarray}}
\newcommand{\eea}{\end{eqnarray}}
\newcommand{\al}{\alpha}
\newcommand{\bt}{\beta}
\newcommand{\gm}{\gamma}
\newcommand{\Gm}{\Gamma}
\newcommand{\dl}{\delta}

\newcommand{\epsv}{\varepsilon}
\newcommand{\zt}{\zeta}

\newcommand{\kp}{\kappa}
\newcommand{\lm}{\lambda}
\newcommand{\ks}{\xi}
\newcommand{\rh}{\rho}

\newcommand{\ups}{\upsilon}

\newcommand{\Om}{\Omega}
\newcommand{\fdot}{\mbox{\boldmath $\cdot$}}
\newcommand{\rarrow}{\rightarrow}

\newcommand{\nn}{\nonumber}

\begin{document}

\title{Geodesic Motions Versus Hydrodynamic Flows in a Gravitating 
Perfect Fluid: Dynamical Equivalence and Consequences}

\author{K. Kleidis$^1$ and N. K. Spyrou$^2$\\
{\small Astronomy Department, Aristoteleion University of Thessaloniki,} \\
{\small 54006 Thessaloniki, Macedonia, Greece (Hellas)} \\
{\small e-mail address: $^1$kleidis@helios.astro.auth.gr , 
$^2$spyrou@helios.astro.auth.gr} }

\maketitle

\begin{abstract}

Stimulated by the methods applied for the observational determination of 
masses in the central regions of the AGNs, we examine the conditions under 
which, in the interior of a gravitating perfect fluid source, the geodesic 
motions and the general relativistic hydrodynamic flows are {\bf dynamically 
equivalent} to each other. Dynamical equivalence rests on the functional 
similarity between the corresponding (covariantly expressed) differential 
equations of motion and is obtained by conformal transformations. In this 
case, the spaces of the solutions of these two kinds of motion are {\bf 
isomorphic}. In other words, given a solution to the problem {\em hydrodynamic 
flow in a gravitating perfect fluid}, one can always construct a solution 
formally equivalent to the problem {\em geodesic motion of a fluid element} 
and vice versa. Accordingly, we show that, the {\em observationally determined 
nuclear mass} of the AGNs is being {\bf overestimated} with respect to the real, 
physical one. We evaluate the corresponding {\bf mass-excess} and show that it 
is not always negligible with respect to the mass of the central dark object, 
while, under circumstances, can be even larger than the rest-mass of the 
circumnuclear gas involved.

\end{abstract}

\section{Introduction}

It is generally believed that active galaxies are powered by the presence of 
central, massive black holes (Rees 1984; Blandford \& Rees 1992). Many galaxies 
may have gone through an active phase and, therefore, massive black holes could 
be common in any {\em active galactic nucleus} (AGN) phenomenon (Holt et al. 
1992; Antonucci 1993; Urry \& Padovani 1995). Moreover, black holes are 
expected to be present in the centers of many quiescent galaxies as well 
(Chokshi \& Turner 1992). However, direct dynamical evidence for the presence 
of black holes in the central regions of individual galaxies is scarce. Stellar 
kinematical studies have provided tentative evidence for black holes only in 
a handfull of nearby galaxies, because of difficulties in spatial resolution 
and the lack of knowledge on the exact shape of the stellar orbits (Kormendy 
\& Richstone 1995; Ho 1998). Based on stellar dynamics, the mass concentrations 
in the central regions of these galaxies are determined through Doppler-shift 
measurements involving Keplerian motions. Accordingly, the velocity dispersion 
$\ups (r)$ of various sources of radiation located at a distance $r$ from the 
center, is usually assumed to follow the standard virial-theorem equation for 
circular geodesic motion (Peebles 1972; Quinlan et al. 1995)
\be
\ups^2 (r) = G \: {M(r) \over r}
\ee
where $M(r)$ is the corresponding distribution of mass inside the radius $r$ 
and $G$ is the Newton's gravitational constant. In fact, it was this law which 
led Sargent et al. (1978) to estimate, for the first time, the mass of the 
central black hole in M87, to be of the order $5 \times 10^9$ $M_{\odot}$.

On the other hand, recent observational data indicate that, in most of the AGNs, 
there exist gas clouds surrounding the central dark object and the associated 
{\em accretion disk}, on a variety of scales from a tenth of a parsec to a few 
hundreds parsecs (see e.g. Holt et al. 1992; Urry \& Padovani 1995). These 
clouds are assumed to form a geometrically and optically thick torus (or 
warped disc), which absorbs much optical UV radiation and soft X-rays 
(Guilbert \& Rees 1988; Rees 1995). Since 1993, many indications appeared in 
favour of the above idea, e.g. by HST imaging of gas and dust in NGC 4261 
(Jaffe et al. 1993), by the mapping in nearby AGNs of the thermal molecular CO 
emission in Centaurs A (Rydbeck et al. 1993), or by the high-density tracer HCN 
in NGC 1068 (Tacconi et al. 1994) and M51 (Kohno et al. 1996). This {\em 
circumnuclear gas} seems to exist in either the molecular or the atomic phase, 
the latter being due to hard X-ray sources in the centers of most radio-loud 
AGNs (Krolik \& Begelman 1988; Maloney et al. 1994). In fact, to date the most 
dramatic evidence for the existence of a supermassive black hole comes from the 
VLBI imaging of molecular $(H_2 O)$ {\em masers} in the active galaxy NGC 4258 
(Greenhill et al. 1995; Miyoshi et al. 1995). This imaging, produced by 
Doppler-shift measurements assuming Keplerian motion of the masering source, 
has allowed a quite accurate estimate of the nuclear central mass, which has 
been found to be a $3.6 \times 10^7$ $M_{\odot}$ supermassive dark object, 
within 0.13 parsecs. {\bf However, this evidence (as well as many of the 
previous ones) comes primarily from the motions of gas streams rather than 
stars}. The gas is also subject to non-gravitational forces and may not follow 
ballistic trajectories. Indeed, it has been recently announced (Greenhill \& 
Gwinn 1997) that the rotation curve of the active galaxy NGC 1068 (in which, 
maser emission has also been detected) is sub-Keplerian, suggesting that 
a non-gravitational component is also present in the maser emission. Therefore, 
as regards the measurement data relevant to the cental dark objects in the AGNs, 
there is some inherent ambiguity (Rees 1995).

In fact, the determination of the central nuclear mass is based on the 
Doppler-shifted radiation received, not from something like a {\em 
test-particle} star, but from the {\em extented} masering fluid sources. 
These sources move in the gravitational field of a (more or less) continuous 
gravitating source, such as the circumnuclear gas and the accretion disk, and 
from a dynamical point of view, the assumption that the motion of the masering 
source is Keplerian (or, in general, {\em geodesic}), seems to be {\em 
inaccurate}. On the other hand, a fluid {\em volume element} can be considered 
as more realistic constituent of a fluid source and hence, its use in a 
continuous medium should be physically preferable than that of a theoretical, 
ideal test-particle (Spyrou 1997a; 1997b; 1997c). So, the question is raised 
as to {\em whether the hydrodynamical flow of a fluid volume element in a 
continuous gravitating source can be approximated or not to the motion of 
a test-particle (namely the geodesic motion) in the same source}.

Motivated by the above considerations, in the present article we examine 
the conditions under which the hydrodynamic flows in the interior of a 
bounded self-gravitating perfect fluid source may become {\bf dynamically 
equivalent} to general-relativistic geodesic motions in this source. It is 
clear that, generally, in the interior of a continuous perfect fluid source 
the hydrodynamic flow differs from the geodesic motion. In this case, 
dynamical equivalence rests in the fact of the functional similarity between 
the corresponding (covariantly expressed) differential equations of motion 
(Spyrou 1999). Accordingly, the spaces of the solutions of these equations 
are {\em isomorphic}. In other words, given a solution to a problem of {\em 
hydrodynamic flow in a gravitating perfect fluid}, one can always construct 
a solution formally equivalent to the problem {\em geodesic motion of a fluid 
element} and vice versa. The problem attacked is kept as general as possible, 
being solved within the context of the exact General Relativity Theory (GRT). 
Therefore, the corresponding results could be useful in many astrophysical 
situations and not just in the context of relativistic galactic dynamics 
and/or cosmology.

\section{Geodesic Motions Versus Hydrodynamic Flows}

The ballistic motion of a test-particle in the gravitational field of a 
bounded, self-gravitating perfect fluid source, takes place along {\em 
geodesic} trajectories, on which the tangent vector of the four-velocity 
$u^i = {d x^i \over d s}$ $(u_i u^i = 1)$ obeys the standard {\em geodesic 
equation} in a coordinate frame $x^i$ (see e.g. Papapetrou 1974)
\be
{D u^i \over d s} = 0 \; \; \; \Leftrightarrow \; \; \; {d u^i \over d s} + 
\Gm_{kl}^i u^k u^l = 0
\ee
where the Latin indices, denoting spacetime coordinates, admit the values 
$0, 1, 2, 3$ and the Greek indices, denoting spatial coordinates, admit the 
values $1, 2, 3$. In Eq. (2), ${D u^i \over d s}$ denotes the {\em covariant 
acceleration} vector, $s$ is the {\em affine parameter} and $\Gm_{kl}^i$ are 
the {\em Christoffel symbols}. On the other hand, the equations of motion of 
the hydrodynamic flow in the interior of the same source, are written in form 
\be
{\cal T}_{\; ; \: j}^{i j} = 0
\ee
where a semi-colon denotes covariant differentiation and ${\cal T}^{ij}$ is the 
stress-energy tensor of the perfect fluid, which is written in the standard form
\be
{\cal T}^{ij} = ({\cal E} + p) u^i u^j - p g^{i j}
\ee
In Eq. (4), $p$ is the isotropic pressure and ${\cal E}$ is the overall 
mass-energy density of the perfect fluid. Both quantities are considered to 
transform like scalars under coordinate mappings (Anderson 1967). In the 
absence of {\em shear} and {\em viscocity}, ${\cal E}$ is decomposed as 
follows (Fock 1959; Chandrasekhar 1965)
\be
{\cal E} = \rh c^2 + \rh \Pi
\ee
where, $\rh c^2 \neq 0$ is the rest-mass energy density of the total number of 
baryons included in the unit volume element, and $\rh \Pi$ is the corresponding 
specific internal energy density, the variations of which are related to the 
compressions or the expansions of the fluid. Combination of Eqs. (3) and (4) 
yields
\be
({\cal E} + p)_{\: , \: j} \: u^i \: u^j + ( {\cal E} + p) \: (u_{\: ; j}^i \: 
u^j + u^i \: u_{\: ; \: j}^j) - p_{\: , \: j} \: g^{i j} = 0
\ee

Now, in order to study a possible dynamical equivalence of the hydrodynamic 
flow of a finite-volume element in the interior of a bounded, self-gravitating 
perfect fluid to the corresponding geodesic motions, first of all we need to 
determine the evolution of the fluid's covariant acceleration. In this case, 
it is convenient to study the flow motion on a hypersurface normal to the 
direction of $u^i$. We do so, by contracting Eq. (6) with the {\em projection 
operator}
\be
h_{ik} = g_{ik} - u_i u_k
\ee
for which, we obviously have
\be
h_{ik} u^i = 0
\ee
Then, provided that ${\cal E} + p \neq 0$, i.e. the total {\em enthalpy} of a 
fluid element is non-vanishing, the combination of Eqs. (6) and (7) yields
\be
u_{k ; \: j} u^j = {p_{, \: j} \over ({\cal E} + p)} \left ( \dl_k^j - u_k u^j 
\right )
\ee
Eq. (9) is just one of the two equations of fluid motion, in which Eq. 
(3) splits in the case of a perfect fluid (Taub 1959). This equation 
is completely analogous to the {\em conservation-of-momentum} equation of 
Newtonian hydrodynamics (Ryan \& Shepley 1975). In a Riemannian space 
$(g_{ij ; s} = 0)$, Eq. (9) is written in the form
\be
g_{kl} u_{; \: j}^l u^j = {p_{, \: j} \over ({\cal E} + p)} \left ( \dl_k^j - 
u_k u^j \right )
\ee
from which, on contracting with $g^{m k}$, we obtain
$$
\dl_l^m \left [ {d u^l \over d s} + \Gm_{j s}^l u^j u^s \right ] = 
g^{mk} {p_{, \: j} \over ({\cal E} + p)} \left ( \dl_k^j - 
u_k u^j \right )
$$
thus, resulting to
\be
{d u^m \over d s} + \Gm_{j s}^m u^j u^s = g^{mk} {p_{, \: j} \over ({\cal E} + 
p)} \left ( \dl_k^j - u_k u^j \right )
\ee
or, finally, 
\be
{d u^m \over d s} + \Gm_{j s}^m u^j u^s = {1 \over {\cal E} + p} \: h^{m j} 
\: p_{, \: j}
\ee
From Eq. (12), it becomes evident that, the two kinds of motion are completely 
(not just dynamically) equivalent, provided that the pressure gradient in the 
three-surface normal to $u^i$ [which appears in rhs of Eq. (12)] is zero. This 
quantity measures the response of a particle to non-gravitational fields and, 
therefore, it is responsible for all deviations of flow lines from geodesics 
(see also Misner et al. 1973; Ryan \& Shepley 1975). 

\section{Dynamical Equivalence Conditions}

By definition, dynamical equivalence between geodesic motions and 
hydrodynamic flows rests in the functional similarity between Eqs. 
(2) and (12). In this case, the spaces of the solutions of these 
equations are isomorphic. In other words, given a solution to a 
problem of {\em hydrodynamic flow in a perfect fluid}, one can always 
construct a solution formally equivalent to the problem {\em geodesic 
motion of a fluid element} and vice versa. If we manage to establish 
dynamical equivalence between Eqs. (2) and (12), then we will be able 
to extrapolate every result obtained on the basis of geodesic motions, 
to the more realistic context of general-relativistic fluid hydrodynamics. 
A direct application may concern observational data from many relativistic 
astrophysical (or cosmological) systems, such as the central cores of the 
AGNs (Kormendy \& Richstone 1995; Rees 1995) or the existence and the nature 
of the {\em dark matter} (Saglia 1996) etc. Dynamical equivalence between 
Eqs. (2) and (12) may be performed in two ways: 

\begin{enumerate}

\item By eliminating the rhs of Eq. (12). A direct way to do so is to assume 
{\em isobaric motions}, i.e. $p = constant$. However, in general, this is not 
always a physically necessary condition. 

\item By transfering the problem to a {\em virtual} self-gravitating perfect 
fluid, the metric of which is {\em conformal} to $g_{kl}$. In terms of this 
fluid, Eq. (12) will be subsequently written in the form of Eq. (2).

\end{enumerate}

The latter, is a very delicate method of obtaining dynamically equivalent 
descriptions in any gravitational field theory and not just GRT (e.g. see 
Spyrou 1976; Whitt 1984; Cotsakis 1993). A conformal transformation (Penrose 
1964; Hawking \& Ellis 1973) shrinks or stretches the entire manifold, 
introducing a change in the scale of all lengths and time, which can be 
different for the various world points, but is the same for all spatial 
directions and time at a given point. A conformal transformation of a 
metric is usually described by
\be
{\tilde{g}}_{kl} = \Om^2 (x^i) \: g_{kl}
\ee
for some continuous, non-vanishing, finite, real function $\Om (x^i)$. 
Accordingly, in what follows, we assume that there exists a virtual 
gravitating perfect fluid, producing a {\bf new} metric tensor 
$({\tilde{g}}_{kl})$, in terms of which we may write Eq. (12) in the form 
of Eq. (2), namely
\be
{d {\tilde{u}}^m \over d \tilde{s}} + {\tilde{\Gm}}_{kl}^m {\tilde{u}}^k 
{\tilde{u}}^l = 0.
\ee

Next, we shall try to determine the relation between these two fluids. To begin 
with, we note that when Eq. (13) is applied, the following transformations hold 
(Hawking \& Ellis 1973)
\bea
{\tilde{\Gm}}_{kl}^m & = & \Gm_{kl}^m \: + \: {1 \over \Om} \: \left ( 
\dl_k^m \Om_{, \: l} + \dl_l^m \Om_{, \: k} - g_{kl} g^{ms} \Om_{, \: s} 
\right ) \nn \\
{d {\tilde{u}}^m \over d \tilde{s}} & = & {1 \over \Om} \: {\partial \over 
\partial x^n} \left ( {1 \over \Om} \right ) \: u^m u^n \: + \: {1 \over \Om^2} 
\: {d u^m \over d s} 
\eea

With the aid of Eqs. (15), the geodesic equation (14) may be decomposed in 
terms of the original metric tensor $(g_{kl})$, as follows
\be
{d u^m \over d s} + \Gm_{kl}^m u^k u^l = h^{mn} \: {\partial \over \partial 
x^n} (ln \Om)
\ee
Now, in view of the above definitions, dynamical equivalence between the two 
kinds of motion under consideration, implies 
\be
h^{m n} \: {\partial \over \partial x^n} (ln \Om) = {1 \over {\cal E} + p} \: 
h^{m n} \: p_{, \: n}
\ee

To solve Eq. (17) with respect to $\Om$ is a very difficult task, especially 
when no further conditions are imposed. Nevertheless, there exists a quite 
general physical condition which may help us to reach at some (at least) 
particular solutions. It is the {\em equilibrium hydrodynamics hypothesis}, 
i.e. the constancy of {\em entropy} $({\cal S})$ along the original fluid's 
flow lines (Taub 1959; Misner et al. 1973)
\be
{\cal S}_{, \: m} \: u^m = 0
\ee
In fact, in relativistic astrophysics and cosmology we usually assume ${\cal S}$ 
to be a group invariant (constant in space) and therefore ${\cal S} = const.$ 
(Ryan \& Shepley 1975). Hence, in what follows we consider that the flow motion 
is {\em isentropic}. 

Adiabaticity of the hydrodynamical motions could be physically necessary, 
because then both the thermodynamic and the matter content (the number of 
baryons) of a finite-volume element of perfect fluid remain constant during 
its motion (Chandrasekhar 1983). Then, the general-relativistic analogue of 
the {\em first law of thermodynamics} is valid with respect to the original 
fluid, for $d{\cal S} = 0$. In the case of a perfect fluid source, this law 
may be expressed in the form
\be
\Pi_{, \: m} + p \left ({1 \over \rh} \right )_{, \: m} = 0 
\ee
or, more conveniently,
\be
p_{, \: m} - \rh c^2 \: \left ({{\cal E} + p \over \rh c^2} \right )_{, \: m} 
= 0 
\ee
Inserting Eq. (20) into Eq. (17), we obtain
\be
h^{m n} \: {\partial \over \partial x^n} \: \left [ ln  \left ( \Om \right ) 
\: - \: ln \left ( { {\cal E} + p  \over \rh c^2} \right ) \right ] = 0
\ee
In Eq. (21), we perform the substitution
\be
\Om (x^k) \: = \: { {\cal E} + p \over \rh c^2} \: e^{\Phi (x^k)} 
\ee
Then, the new metric $({\tilde{g}}_{m n})$, in terms of the original one 
$(g_{m n})$, is written in the form
\be
{\tilde{g}}_{m n} \: = \: \left ( { {\cal E} + p \over \rh c^2} \right )^2 \: 
e^{2 \Phi (x^k)} \: g_{m n}
\ee
and the dimensionless, real scalar function $\Phi (x^k)$ satisfies the {\em 
compatibility condition}
\be
h^{m n} \: \Phi_{, \: n} = 0
\ee
Eq. (24) has a clear physical meaning: The projection of the vector $\Phi_{, 
\: n}$ on a three-dimensional hypersurface normal to $u_n$ is zero. By virtue 
of Eq. (8), this condition implies that $\Phi_{, \: n}$ is always {\em 
collinear} to $u_n$. Accordingly, the {\bf general solution} to Eq. (24) may 
be obtained in terms of the differential equation
\be
\Phi_{, \: n} = \zt (x^m) \: u_n
\ee
(in connection see Anderson 1967; Ryan \& Shepley 1975) where $\zt (x^m)$ is 
an arbitrary scalar function with dimensions of inverse length. The general 
solution to Eq. (25), may be expressed in the form
\be
\Phi (x^m) = \int \zt (x^m) \: u_n \: dx^n + \ks (x^r)
\ee
where no summation is assumed over $n$, and $\ks (x^r)$ is a dimensionless 
scalar function of the coordinates $x^r$ $(r \neq n)$. Obviously, the exact 
form of the general solution to Eq. (24) remains undetermined. This is not 
an unexpected result, since Eq. (24) is a system of four partial differential 
equations involving five unknown quantities, namely, the four components of 
$u^i (x^n)$ and $\Phi (x^n)$. In this case, supplementary conditions should 
be imposed, for someone to deal with a well-defined problem. 

Nevertheless, there exists one {\bf particular solution} to Eq. (24) which is 
of great theoretical interest, namely
\be
\Phi (x^n) = {\cal C} = const.
\ee
In this case, the combination of Eqs. (23) and (27) relates the physical 
metric tensor $g_{m n}$, due to the gravitational field in the interior 
of a real perfect fluid source (in which the motions follow the laws of 
relativistic hydrodynamics), to the correponding metric ${\tilde {g}}_{m 
n}$ of a virtual fluid (in which the original hydrodynamic motions become 
geodesic). According to Eq. (23), the functional form of the metric 
attributed to the new gravitational source includes the overall enthalpy 
content $({\cal E} + p)$ within a {\em specific volume} $({1 \over \rh})$ 
of the original one ({\em specific enthalpy}). 

The family of conformal transformations (13) [or (23)] represents an {\em 
algebraic group} over the space of the Riemannian metrics. In this case, 
it is evident that the corresponding {\em operation} among the various group 
elements is the standard multiplication between the real, finite, non-zero, 
scalar functions $\Om (x^n)$. There exists an {\em identity} element 
$(\Om = 1)$, as well as an {\em inverse} one $(\Om^{-1})$, such that $\Om^{-1} 
\cdot \Om = 1$. As regards the identity element, it represents the degenerate 
case, in which the associated metrics (the real and the virtual) are {\bf 
identical}, ${\tilde{g}}_{m n} = g_{mn}$. In our problem, this corresponds 
to an {\em isobaric flow} of the original fluid [e.g. see Eq. (12)]. For $p 
= p_0 = const.$, the first law of thermodynamics [Eq. (19)] is reduced to 
\be
\Pi + {p_0 \over \rh} = const. 
\ee
and the identity transformation, through Eqs. (23) and (27), implies
\be
{\cal C} = - \: ln \left [1 + {1 \over c^2} ( \Pi + {p_0 \over \rh}) \right ] 
= const.
\ee
Therefore, if the adiabatic flow of a finite-volume element in the interior 
of a bounded, self-gravitating perfect fluid is (in addition) isobaric, then 
the corresponding metric tensor is {\bf conformally invariant} under the 
group of transformations (23), and the hydrodynamic flows are completely 
equivalent to geodesic motions. 

In what follows, we shall assume the validity of Eq. (27) and, unless 
otherwise stated, we shall ignore the constant factor $e^{2 {\cal C}}$ 
arising in the formula for $\Om$ from the combination of Eqs. (23) and 
(27). This normalization may be performed through a redefinition of the 
proper length in the original metric, as $s \rarrow \lm = s \: e^{- {\cal 
C}}$. In concluding, the appropriate conformal transformation which 
guarantees the dynamical equivalence between hydrodynamic flows and 
geodesic motions in a gravitating perfect fluid, is 
\be
{\tilde{g}}_{m n} \: = \: \left ( { {\cal E} + p \over \rh c^2} \right )^2  
\: g_{m n} 
\ee
provided that the following major assumptions are valid:

\begin{enumerate} 

\item We have considered a gravitating perfect fluid, i.e. isotropic in 
pressure, with no shear and viscocity.

\item In this source, the hydrodynamic flow motions are adiabatic, i.e. 
${\cal S}$ is constant along the fluid flow lines.

\item The gravitational field in the interior of this fluid corresponds to 
a Riemannian spacetime, i.e. $g_{\: kl \: ; \: m} = 0$ and $det \vert g_{kl} 
\vert \neq 0$.

\item The corresponding metric tensor is a solution to the Einstein field 
equations, through ${\cal T}_{\: ; \: l}^{kl} = 0$.

\end{enumerate}

\section{Reduction to the Newtonian Limit}

In many cases of particular astrophysical interest, the corresponding {\em 
weak field-limit} results seem to be more appropriate. For example, it has 
already been shown (Spyrou \& Varvoglis 1982) that, in the context of 
relativistic galactic dynamics and astrophysics, the use of the first 
post-Newtonian approximation is more than enough in most of the cases. In 
fact, for astrophysically interesting sources with non-negligible relativistic 
characteristics (such as e.g. the nuclei of the giant elliptical galaxies) 
the small post-Newtonian expansion parameter (for the solar system being 
$\approx 10^{-6}$) is still very small $( \approx 10^{-3} \ll 1)$. Accordingly, 
we will examine how the above results, obtained within the context of the exact 
GRT, may affect on the corresponding Newtonian limit. In this limit, as regards 
the metric tensor of the original gravitating perfect fluid, we have (Schiff 
1960; Chandrasekhar 1965)
\bea
g_{0 0} & = & 1 - 2 {U \over c^2} + O_4 \nn \\
g_{0 \al} & = & O_3 \nn \\
g_{\al \bt} & = & - \dl_{\al \bt} + O_2
\eea
where the symbol $O_n$ denotes terms of order $n$  in the small parameter 
${\epsv \over c} \ll 1$, with 
\be
\epsv^2 = max \: (\ups^2 , U , \Pi , {p \over \rh} )
\ee
with $\ups$ denoting the measure of the spatial three-velocity vector and $U$ 
the gravitational potential, related to the mass density $\rh$ through the 
standard Poisson equation
\be
\nabla^2 U = - 4 \pi G \: \rh
\ee

According to Eq. (30), the components of the metric tensor attributed to the 
virtual perfect fluid, may be written in terms of the corresponding components 
of the original one (31), as follows
\bea
{\tilde{g}}_{0 0} & = & \left ( { {\cal E} + p \over \rh c^2} \right )^2 \: 
\left ( 1 - 2 {U \over c^2} + O_4 \right ) \nn \\
& = & \left ( 1 + 2 \: {1 \over c^2} \: \left [ \Pi + {p \over \rh} \right ] 
+ O_4 \right ) \: \left ( 1 - 2 {U \over c^2} + O_4 \right ) \nn \\
& = & 1 - 2 \: {{\tilde{U}} \over c^2} + O_4
\eea
In this case, from Eq. (34) we observe that, ${\tilde{g}}_{00}$ is formally 
equivalent to $g_{00}$, provided that $U$ is replaced by the {\em virtual 
gravitational potential}
\be
{\tilde{U}} = U - \left (\Pi + {p \over \rh} \right )
\ee
Eq. (35) is identical to the corresponding Newtonian result of Spyrou (1997a; 
1997b; 1997c). In the same fashion, we obtain
\be
{\tilde{g}}_{0 \al} = O_3 = g_{0 \al}
\ee
and
\bea
{\tilde{g}}_{\al \bt} & = & \left ( - \dl_{\al \bt} + O_2 \right ) \: \left [ 
1 + 2 \: {1 \over c^2} \: \left ( \Pi + {p \over \rh} \right ) + O_4 \right ] 
\nn \\
& = & - \dl_{\al \bt} + O_2 \nn \\
& = & g_{\al \bt}
\eea
The set of Eqs. (34) - (37) represents the appropriate weak field-limit 
transformation, relating the conformal metrics of the real and the virtual 
perfect fluid, under the action of the group (30). In analogy to Eq. (33), 
we may define the {\em virtual rest-mass density} $({\tilde{\rh}})$, which 
produces the corresponding gravitational potential ${\tilde{U}}$, through 
\be
\nabla^2 {\tilde{U}} = - 4 \pi G \: {\tilde{\rh}}
\ee
which, in view of Eq. (35), yields
\be
{\tilde{\rh}} = \rh + {1 \over 4 \pi G} \: \nabla^2 ( \Pi + {p \over \rh} ).
\ee

\section{The Conformal Stress-Energy Tensor}

The conformal transformation (30) affects both the geometry and the physical 
quantities relevant to the matter content of the gravitating source (the 
stress-energy tensor of the perfect fluid), in a way that the dynamical 
description of gravity in the virtual case (the functional form of the 
Einstein field equations) remains invariant. In this respect, we consider 
that ${\tilde{g}}_{m n}$ is a solution of the {\em conformal Einstein field 
equations}
\be
{\tilde{\cal R}}_{ml} - {1 \over 2} {\tilde{g}}_{ml} {\tilde{\cal R}} = 
- \kp {\tilde{\cal T}}_{ml}
\ee
where also
$$
\kp = {8 \pi G \over c^4}
$$
By virtue of Eqs. (40), it is evident that the corresponding {\em conformal 
stress-energy tensor} $({\tilde{\cal T}}_{ml})$ satisfies the conservation law 
${\tilde{\cal T}}^{ml}_{\; ; \: l} = 0$, where now the covariant derivative 
is taken with respect to the conformal metric, ${\tilde{g}}_{m l}$. In this 
case, we may deduce the exact form of the stress-energy tensor, responsible 
for the gravitational field of the virtual fluid, in terms of the corresponding 
tensor of the original source.

In four dimensions, the influence of a conformal transformation of the form 
(13) on the dynamical quantities describing the gravitational field (the 
Riemann tensor and its contractions) is (Hawking \& Ellis 1973)
\be
{\tilde{\cal R}}_{ml} = {\cal R}_{ml} - 2 \Om \: \left ( {1 \over \Om} 
\right )_{; \: m l} + {1 \over 2} g_{ml} {1 \over \Om^2} \: \Box \left 
( \Om^2 \right )
\ee
and 
\be
{\tilde{\cal R}} = {1 \over \Om^2} {\cal R} + 6 {1 \over \Om^3} \Box \Om
\ee
where the d' Alembert operator $(\Box)$, with respect to the original metric, 
is given by
$$
\Box \Om = {1 \over \sqrt{-g}} {\partial \over \partial x^m} \: \left ( 
\sqrt{-g} \: g^{mn} \: \Om_{, \: n} \right ).
$$
With the aid of transformations (41) and (42), Eq. (40) reduces to 
\bea
{\tilde{\cal T}}_{ml} & = & {\cal T}_{ml} + {2 \over \kp} \: g_{ml} \: 
{\Box \Om \over \Om} - {2 \over \kp} \: {1 \over \Om} \: \Om_{; \: m l} \nn \\
& + &  {4 \over \kp} \: {1 \over \Om^2} \: \Om_{, \: m} \Om_{, \: l} - 
{1 \over \kp} \: g_{ml} \: {1 \over \Om^2} \: g^{ns} \Om_{, \: n} \Om_{, \: s}
\eea
and therefore, the trace of ${\tilde{\cal T}}_{ml}$ is written in the form
\be
{\tilde{\cal T}} = {\tilde{g}}^{ml} \: {\tilde{\cal T}}_{ml} = {1 \over \Om^2} 
\: \left [ {\cal T} + {3 c^4 \over 4 \pi G} \: {\Box \Om \over \Om} \right ]
\ee
Clearly, Eqs. (43) and (44) hold for an arbitrary form of both 
${\tilde{\cal T}}_{ml}$ and ${\cal T}_{ml}$. However, in connection to what 
previously stated, we may assume that the conformal stress-energy tensor can 
be written in the form of a perfect fluid, namely, 
\be
{\tilde{\cal T}}_{ml} = \left ( {\tilde{\cal E}} + {\tilde{p}} 
\right ) {\tilde{u}}_m {\tilde{u}}_l - {\tilde{p}} \: {\tilde{g}}_{ml}
\ee
for which we have
\be
{\tilde{\cal T}} = {\tilde{\cal E}} - 3 \: {\tilde{p}}
\ee
where, ${\tilde{\cal E}}$ and ${\tilde{p}}$ are the corresponding energy density 
and pressure. In this case, Eq. (45) results to
\be
{\tilde{\cal E}} - 3 {\tilde{p}} = {1 \over \Om^2} \: \left [ {\cal E} - 3 p 
+ {3 c^4 \over 4 \pi G} \: {\Box \Om \over \Om} \right ]
\ee
By virtue of Eq. (47) we perform the {\em ansatz} that, under the action of 
the conformal group (30), both ${\cal E}$ and $p$ transform as follows
\bea
{\tilde{\cal E}} & = & {1 \over \Om^2} \: \left [ {\cal E} + a \: {c^4 \over 4 
\pi G} \: {\Box \Om \over \Om} \right ] \\
{\tilde{p}} & = & {1 \over \Om^2} \: \left [ p - ( {3 - a \over 3} ) \: {c^4 
\over 4 \pi G} \: {\Box \Om \over \Om} \right ]
\eea
where $a$ is a numerical constant, the exact value of which can be determined 
in terms of the corresponding weak field-limit results [Eq. (39)] (see also 
Spyrou 1999). In analogy to Eq. (5), we assume that ${\tilde{\cal E}}$ may be 
decomposed as follows
\be
{\tilde{\cal E}} = {\tilde{\rh}} c^2 \: (1 + {{\tilde{\Pi}} \over c^2})
\ee
where ${\tilde{\rh}} {\tilde{\Pi}}$ corresponds to the specific internal energy 
density, related with the contractions or the expansions of the virtual fluid. 
Then, in the weak field-limit, Eq. (48) is reduced to
\bea
{\tilde{\rh}} \: ( 1 + {{\tilde{\Pi}} \over c^2} ) & = & \rh \: 
\left [ 1 - {1 \over c^2} ( \Pi + 2 {p \over \rh} ) \right ] + \nn \\
& + & a \: {1 \over 4 \pi G} \: \left [ 1 - {3 \over c^2} ( \Pi + 
{p \over \rh} ) \right ] \: \left ( {1 \over c^2} \: {\partial^2 \over 
\partial t^2} - \nabla^2 \right ) \: \left [\Pi + {p \over \rh} \right ]
\eea
which, in the stationary case, to lowest order in ${1 \over c^2}$ yields
\be  
{\tilde{\rh}} = \rh - a \: {1 \over 4 \pi G} \: \nabla^2 \: 
\left (\Pi + {p \over \rh} \right )
\ee
Now, by comparison of Eqs. (39) and (52), we obtain $a = -1$. Hence, Eqs. 
(48) and (49) are finally written in the form
\bea
{\tilde{\cal E}} & = & {1 \over \Om^2} \: \left ( {\cal E} - {c^4 \over 4 \pi G} 
\: {\Box \Om \over \Om} \right ) \\
{\tilde{p}} & = & {1 \over \Om^2} \: \left ( p - {c^4 \over 3 \pi G} \: 
{\Box \Om \over \Om} \right )
\eea
obviously satisfying Eqs. (47) - (49). As regards a self-gravitating perfect 
fluid, Eqs. (53) and (54) represent the transformation law of the corresponding 
energy density and pressure, under the action of the conformal group (30). It 
is worth noting that, according to Eq. (54), {\em the pressure of the virtual 
perfect fluid may not be constant (or vanish), although the motions in its 
interior are geodesics}. This result is a distinguishing feature of the present 
analysis, with respect to what we have encountered so far in relativistic 
astrophysics and cosmology, namely, that the isobaric flows are geodesics (see 
e.g. Weinberg 1972; Misner et al. 1973; Papapetrou 1974; Ryan \& Shepley 1975; 
Narlikar 1983) and so, it justifies the use of the term {\em "virtual"}, in 
denoting the matter content responsible for the conformal metric.

In relation to the measurement data relevant to what we observe and determine 
observationally in the central regions of the AGNs, the theoretical result (53) 
has a clear physical interpretation: ${\tilde{\cal E}}$ is what we actually 
measure by assuming geodesic motions, while ${\cal E}$ is the real physical 
quantity associated to the fluid matter content existing in those regions. 
In this respect, the quantity
\be
\rh_i = - {c^2 \over 4 \pi G} \: {\Box \Om \over \Om}
\ee
may be identified as an {\bf extra} inertial energy density, associated to 
the contributions from the {\em internal physical characteristics (internal 
motions, pressure, thermodynamic content)} of the original fluid, to the 
measured quantities. In fact, $\rh_i$ describes the way the various physical 
characteristics of the source (beyond its rest-mass density) act as 
additional gravitational sources, thus affecting the geodesic motions 
(Spyrou \& Dionysiou 1973; Spyrou 1997a). In this case, Eq. (53) is written 
in the form
\be
{\tilde{\cal E}} = {1 \over \Om^2} \: \left ( {\cal E} + \rh_i c^2 
\right )
\ee
and the question arises as to whether the observationally measured energy 
${\tilde{\cal E}}$ is being overestimated or not, with respect to the 
physical quantity ${\cal E}$. 

According to Eqs. (55) and (56), the answer to the above question depends 
on the {\em sign} of $\Box \Om$, which can be determined by taking into 
account the so called {\em weak energy condition}. It is known (Hawking 
\& Penrose 1970) that, for any physical system (as it is the case for the 
original perfect fluid, described by ${\cal T}_{m l}$), the weak energy 
condition is valid, i.e. the locally measured energy is non-negative
\be
{\cal T}_{m l} \: u^m u^l = {\cal E} \geq 0
\ee
Admiting that the virtual fluid also represents a physical system, we 
demand that a condition similar to Eq. (57) is valid with respect to 
${\tilde{\cal T}}_{m l}$ as well, namely,
\be
{\tilde{\cal T}}_{m l} \: {\tilde{u}}^m {\tilde{u}}^l = {\tilde{\cal E}} \geq 0.
\ee
With the aid of Eq. (53), Eq. (58) is decomposed in terms of ${\cal E}$, 
as follows
\be
{\cal E} \geq {c^4 \over 4 \pi G} \: {\Box \Om \over \Om}
\ee
Now, by virtue of Eq. (57), the condition (59) is valid {\bf for all} ${\cal E} 
\geq 0$, provided that
\be
\Box \Om \leq 0
\ee
Therefore, we conclude that $\rh_i \geq 0$ and according to Eq. (56), 
the observationally determined energy $({\tilde{\cal E}})$ {\bf is being 
overestimated} as compared to the real, physical one $({\cal E})$. The 
equality in Eq. (60) corresponds to the identity transformation case, 
representing an isobaric flow of the original fluid, for which, from Eqs. 
(20) and (22), we readily obtain $\Box \Om = 0$. 

\section{Some Astrophysical Applications}

Recently, it has been proposed that the centers of the AGNs contain large 
amounts of gas, in the form of warped disks or tori, which absorbs much 
of the non-thermal power generated in the central core (probably due to 
accretion phenomena) and reradiate it (Antonucci \& Miller 1985; Guilbert 
\& Rees 1988; Holt et al. 1992; Jaffe et al. 1993; Rydberg et al. 1993; 
Tacconi et al. 1994; Kohno et al. 1996; Greenhill \& Gwinn 1997). Direct 
mapping of these gaseous structures can be possible with high spatial 
resolution spectroscopic observations of {\em water masers} (Urry \& 
Padovani 1995). Indeed, VLBI imaging of water masers in NGC 4258 has 
provided a firm evidence for rotating molecular gas in a region less 
than $0.25$ $pc$ from the center of the galaxy (Greenhill et al. 1995; 
Miyoshi et al. 1995). Assuming Keplerian motion of the masering source, 
this imaging has allowed a quite accurate estimate of the central nuclear 
mass, which has been found to be a $3.6 \times 10^7 \: M_{\odot}$ supermassive 
dark object, within $0.13$ $pc$. However, combining these observations with 
the theoretical results obtained in the present article, it becomes evident 
that assuming Keplerian (geodesic) motions in the central region of the AGNs 
corresponds to work within the context of the virtual fluid. In this case, 
the {\em virtual results} can be extrapolated to the real, physical ones 
(based on hydrodynamic motions) through Eqs. (30), (53) and (54). 

As a direct application, we will calculate the {\bf mass-excess} $m_i$, 
arising in the determination of the central nuclear mass of the active galaxy 
NGC 4258, due to the assumption of Keplerian motion of the masering source. In 
fact, $m_i$ represents the contribution of the fluid's internal characteristics 
$(\rh_i)$ to the measured quantities. The corresponding calculations will be 
carried out in the weak field limit, since the various measurements based on 
observational data relevant to the central region of the AGNs, are usually 
derived by methods of Newtonian gravity. In this limit, the {\em superposition 
principle} is valid and therefore, each gravitating source (the black hole 
and/or the circumnuclear gas) is expected to contribute separately. For the 
sake of simplicity we shall assume that the nuclear region is {\em spherically 
symmetric}, so that all the physical quantities describing the galaxy's interior 
are functions of the radial coordinate $r$. We consider an idealized model, 
according to which the circumnuclear gas surrounding the central dark object 
can be represented by a stationary perfect fluid which undergoes adiabatic, 
hydrodynamical flow under the influence of its own gravitational field, 
plus the gravitational field of the black hole (Novikov \& Thorne 1973). 
In this case, dynamical equivalence between hydrodynamic flow and the 
Keplerian motion of a fluid's volume element implies that Eq. (35) holds, 
provided that the original gravitational potential $U$ is now replaced by 
$U_{BH} + U_G$, where
\be
U_{BH} = G \: {M_{BH} \over r}
\ee
is the gravitational potential due to a (Schwarzschild) black hole of mass 
$M_{BH}$, and $U_G$ is the corresponding potential of the self-gravitating 
gaseous matter in this region. Now, using Poisson's equation, Eq. (39) takes 
on the form
\be
{\tilde{\rh}} = \rh + {1 \over 4 \pi G} \: \nabla^2 (\Pi + {p \over \rh}) 
- {1 \over 4 \pi} \: M_{BH} \: \nabla^2 ({1 \over r})
\ee
where $\nabla^2 ({1 \over r}) = - 4 \pi \dl (r)$. The integration of Eq. (62) 
over a spatial volume ${\cal V}$ of linear dimension $0.25$ $pc$, will give us 
the {\em virtual rest-mass} $({\tilde{m}})$ of the central region in NGC 4258, 
namely,
\be
{\tilde{m}} = ( m + M_{BH} ) + m_i
\ee
where $m$ is the total rest-mass of the gas included in this region, while 
$m_i$ represents the {\em relative error} introduced in the determination 
of the central mass by ignoring the contribution of the fluid's internal 
characteristics. Therefore, it becomes evident that the measured mass 
${\tilde{m}}$ is being {\em overestimated} with respect to the real, 
physical quantity $m + M_{BH}$.

To determine the integral $m_i = \int_{\cal V} \rh_i d^3 x$, we consider that 
the lower limit of integration corresponds to the radius of the {\em innermost 
stable circular geodesic} 
\be
r_0 = 3 R_S = {6 G M_{BH} \over c^2}
\ee
We see that the value of $r_0$ is directly proportional to the mass of the 
central black hole. In general, the estimated masses of the galactic nuclear 
dark objects fall in the range $10^7 - 10^9 \: M_{\odot}$ (Kormendy \& Richstone 
1995). Accordingly, the corresponding values of $r_0$ range from $10^{-5}$ 
$pc$ to $10^{-3}$ $pc$ (in connection see Holt et al. 1992). In the present 
article we adopt a {\em mean value} of $10^{-4}$ $pc$. As regards to the 
upper limit of integration, we take $r_{max} = 0.25$ $pc$, which corresponds 
to the outer boundary of the masering annulus in NGC 4258 (Miyoshi et al. 
1995). Moreover, by virtue of the adiabaticity condition (19), we obtain
\be
{1 \over 4 \pi G} \: \nabla^2 (\Pi + {p \over \rh}) = \rh_i = {1 \over 4 
\pi G} \: \nabla \fdot ({1 \over \rh} \: \nabla p)
\ee
In the special case of adiabatic motions, the pressure is related to the 
rest-mass density through the {\em equation of state} 
\be
p = k \: \rh^{\gm}
\ee
where $k > 0$ and $\gm \geq 1$ are constants. Now Eq. (65) takes on the 
form
\be
\rh_i = {k \gm \over 4 \pi G} \: \rh^{\gm - 2} \: \left [ \nabla^2 \rh + 
(\gm - 2) \: {1 \over \rh} \: (\nabla \rh \fdot \nabla \rh ) \right ]
\ee
In particular, we consider that the proper-mass density distribution can 
be described by a Plummer-type function (Maoz 1995) of the form
\be
\rh = \rh_0 \: (1 + x^2)^{- {n \over 2}}
\ee
where $x = {r \over r_0}$, while $\rh_0$ and $n$ are possitive constants. 
In this case, Eq. (67) results to
\be
\rh_i (x) = {k \gm n \over 4 \pi G} \: {\rh_0^{\gm -1} \over r_0^2} \: 
{1 \over (1 + x^2)^{{n \over 2} (\gm - 1) + 2}} \: \left \lbrace [n (\gm - 1) 
+ 1] x^2 \: - \: 1 \right \rbrace
\ee
and the associated mass is written in the form
\be
m_i = {k \gm n \over G} \: r_0 \: \rh_0^{\gm -1} \: \int_{x_0}^{x_{max}} 
{x^2 \: \left \lbrace [n (\gm - 1) + 1] x^2 \: - \: 1 \right \rbrace \over 
(1 + x^2)^{{n \over 2} (\gm - 1) + 2}} \: d x
\ee
where $x_0 = 1$ and $x_{max} = 0.25/10^{-4} = 2500$. The integral on the rhs 
of Eq. (70) can be expressed in terms of elementary functions only if
\be
{n \over 2} \: (\gm - 1) \: + \: 2 = \ell \: : an \: integer
\ee
(Gradshteyn \& Ryzhik 1965) and in fact, on physical grounds ($n > 0$ 
and $\gm \geq 1$), we must have $\ell \geq 2$. The case $\ell = 2$ 
corresponds, for every value of $n$, to an {\em isothermal flow} of the 
perfect fluid, i.e. $\gm = 1$. We shall consider this case separately. For 
$\ell \geq 2$, the general solution to Eq. (70) is written in the form 
(Gradshteyn \& Ryzhik 1965)
\bea
m_i & = & {k \gm n \over G} \: r_0 \: \rh_0^{\gm -1} \: \left [ -  
{2 \ell - 3 \over 2 \ell - 5} \: {x^3 \over (1 + x^2)^{\ell - 1}} - 
{4 (\ell - 1) \over (2 \ell - 3) (2 \ell - 5)} \: {x \over (1 + 
x^2)^{\ell - 1}} \right. \nn \\
& + & {4 (\ell - 1) \over (2 \ell - 1) (2 \ell - 3) (2 \ell - 5)} \: 
\sum_{j = 1}^{\ell - 1} {(2 \ell - 1) ... (2 \ell - 2 j + 1) 
\over (\ell - 1) ... (\ell - j)} \: {1 \over 2^j} \: 
{x \over (1 + x^2)^{\ell - j}} \nn \\
& + & \left. {4 (\ell - 1) \over (2 \ell - 3) (2 \ell - 5)} \: {(2 \ell - 3) !! 
\over (\ell - 1)!} \: {1 \over 2^{\ell - 1}} \: tan^{-1} x 
\right ]_{x_0}^{x_{max}} 
\eea
Clearly, Eq. (72) is too complicated to be useful in astrophysical 
applications. On the other hand, the isothermal case $(\ell = 2)$ 
seems to be more appropriate. Indeed, recent results regarding the 
properties of {\em mass accretion} in NGC 4258 (Neufeld \& Maloney 
1995), indicate that the adiabatic flow of the the gaseous fluid is 
actually isothermal, with a constant {\em speed of sound} $c_s = 
7 \: km/sec$. In this case, $\gm = 1$ and, therefore, Eq. (72) is 
written as
\be
m_i = {k n \over G} \: r_0 \: \left [ x \: ( {x^2 + 2 \over x^2 + 1} ) 
- 2 tan^{-1} x \right ]_{x_0}^{x_{max}}
\ee
which, by virtue of Eq. (64), results to
\be
m_i = (k n) \fdot 1.66 \times 10^{- 17} \; M_{BH}
\ee
However, in the case of isothermal flow, we furthermore have 
\be
k = c_s^2
\ee
and, therefore, as regards NGC 4258, adopting the typical values $c_s = 7 \: 
km/sec$ (Neufeld \& Maloney 1995) and $n = 5$ (Miyoshi et al. 1995) we 
finally obtain
\be
m_i \sim 5 \times 10^{-5} \; M_{BH}
\ee
This very small relative error, is consistent with observations and in 
fact, provides a theoretical explanation for the almost perfect Keplerian 
rotation-curve observed for the gas, in the central region of this particular 
galaxy (Greenhill et al. 1995; Maoz 1995; Miyoshi et al. 1995). Nevertheless, 
had we applied the same model in NGC 1068, where the outer radius of the 
masering source is $r_{max} = 1$ $pc$ (Greenhill \& Gwinn 1997), we would have 
obtained
\be
m_i \sim 2 \times 10^{-4} \; M_{BH}
\ee
while, in the extreme case of NGC 4261, where (non-masering) circumnuclear 
gas and dust appears to extend up to $r_{max} \sim 150$ $pc$ from the center 
(Jaffe et al. 1993), we would have accordingly obtained
\be
m_i \sim 3 \times 10^{-2} \; M_{BH}
\ee
Therefore, $m_i$ {\bf is not always negligible compared to the mass of the 
central dark object} and it can account from a few hundredths of thousanths 
to several hundredths of $M_{BH}$, depending on the linear dimensions of the 
circumnuclear gas observed in the AGNs.

On the other hand, the assumption of Keplerian motions in the cental region 
of the AGNs is compatible with the existence of circumnuclear gas around a 
black hole, only if the central dark object contains at least 98\% of the 
galactic nuclear mass (Greenhill et al. 1995). In fact, calculations based 
on Newtonian dynamics indicate that the total rest-mass of the gas is $m 
\sim {1 \over 50} \: M_{BH}$ (Maoz 1995). In this case, we may also calculate 
the ratio ${m_i \over m}$ for the particular galaxies considered above. 
Accordingly, for NGC 4258, we find
\be
m_i = 2.5 \times 10^{-3} \; m
\ee
while for NGC 1068 and NGC 4261 we obtain $m_i \sim 10^{-2} \: m$ and $m_i 
\sim 1.5 \: m$, respectively. Therefore, $m_i$ {\bf can be comparable to or 
even larger than the total rest-mass of the circumnuclear gas involved}.

At the outcome, we have to point out that the above idealized scheme is 
rather naive in realistic situations. In fact, it is expected that in the 
central regions of the AGNs the flow motion is not adiabatic at all. 
Significant energy losses may occur, due to thermal bremsstrahlung, 
synchrotron radiation and/or radiative transfer (Novikov \& Thorne 1973), 
leading to a loss of angular momentum, so that, ultimately, the matter 
surrounding the central dark object will end up into the hole (Shakura 
\& Sunyaev 1973). It has been found (Neufeld \& Maloney 1995) that, for 
a mass accretion rate of the order $\dot{m} = 7 \times 10^{-5} \: M_{\odot} 
\: yr^{-1}$, material located at a distance $0.25$ $pc$ from the center 
will fall into the hole after $4 \times 10^6 \: yrs$. Therefore, the case 
of a {\em viscous accretion disk} seems to be more appropriate than the 
spherically symmetric perfect fluid distribution, but it will not be 
considered here. 

\section{Discussion and Conclusions}

Although recent developments in galactic dynamics indicate that the center 
of the AGNs contains large amounts of gas (Urri \& Padovani 1995), the 
observational determination of the central nuclear masses is based on the 
assumption of pure geodesic motions (Kormendy \& Richstone 1995). Stimulated 
by this contradiction, we have examined the conditions under which the 
hydrodynamic flow in the interior of a self-gravitating perfect fluid can 
become dynamically equivalent to geodesic motion in the same source. By 
definition, dynamical equivalence rests on the functional similarity between 
the corresponding differential equations of motion. In this case, the spaces 
of the solutions of these two kinds of motion are {\em isomorphic}. In other 
words, given a solution to a problem of hydrodynamic flow in a gravitating 
perfect fluid, one can always construct a solution formally equivalent to 
the problem of geodesic motion in the same source and vice versa.

As regards a self-gravitating perfect fluid with metric $g_{ml}$, dynamical 
equivalence between the two kinds of motion can be achieved in the case of 
{\em adiabatic flow}. Then, we actually transfer the problem to a {\em virtual} 
fluid, with metric ${\tilde{g}}_{ml}$, in which the motions are geodesics. In 
this case, the two metrics are connected by means of a {\em conformal 
transformation} [Eq. (13)], with the identity element representing an 
{\em isobaric flow} of the original fluid.

In general, to determine the functional form of the {\em conformal factor} 
$\Om (x^i)$, involves the solution of the {\em compatibility equation} (24). 
In the present article we have considered the particular solution $\Phi = 
constant$. The resulting conformal transformation affects both the geometry 
and the stress-energy tensor $({\cal T}_{ml})$ of the original fluid, in a 
way that the functional form of the Einstein field equations remains invariant. 
In this respect, we have also determined the form of the {\em virtual 
stress-energy tensor} $({\tilde{\cal T}}_{ml})$ in terms of the original one. 
In comparison to the measurements relevant to the observational data from the 
central region of the AGNs, these two quantities have a clear physical 
interpretation: ${\tilde{\cal T}}_{ml}$ is what we actually measure by assuming 
geodesic motions, while ${\cal T}_{ml}$ corresponds to the "real" stress-energy 
tensor associated to the fluid matter content in those regions. In this case, 
with the aid of the weak energy condition, we have found that the 
observationally determined quantity $({\tilde{\cal T}}_{ml})$ is being 
{\em overestimated} with respect to ${\cal T}_{ml}$.

Admiting an idealized model in which a spherically symmetric perfect fluid 
surrounds a massive black hole, we have applied the previous results in the 
determination of the central nuclear mass, as regards some particular AGNs. 
In the Newtonian limit, the corresponding results indicate that the {\em 
mass-excess} $(m_i)$ arising in the observational determination of the 
central masses by ignoring the contribution of the fluid's internal 
characteristics, is not always negligible compared to the mass of the black 
hole. In fact, it can account from a few hundredths of thousanths (in NGC 
4258) to several hundredths (in NGC 4261) of $M_{BH}$, depending on the 
linear dimension of the circumnuclear gas. On the other hand, we have shown 
that $m_i$ can be either comparable or even larger than the total rest-mass 
of the fluid involved. 

However, the assumption of a (spherically symmetric) perfect fluid can be 
rather naive in realistic situations. In this case, the {\em viscous 
hydromagnetic flow} in the interior of a {\em warped accretion disk} should 
be more appropriate, but the corresponding equations are too complicated for 
astrophysical applications. Furthermore, it can be verified that in the case 
of a {\em magnetized self-gravitating perfect fluid}, the determination of the 
corresponding conformal transformation involves the solution of a compatibility 
condition of the form
\be
h^{mn} \: {\Om_{, \: n} \over \Om} = - \: {e \over m_e c} \: {\cal F}_n^m \: 
u^n
\ee
where, $m_e$ is the mass of the charge $e$ and ${\cal F}_n^m$ is the 
antisymmetric tensor of the electromagnetic field involved. Clearly, Eq. 
(80) is far more complicated than Eq. (24).

Finally, one cannot help asking whether dynamical equivalence between the 
two kinds of motion implies physical equivalence, in general. Clearly, the 
geodesic motion of a test-particle is an approximate description with certain 
limits of validity. The test-particle does not react back to modify the 
original gravitational field. To decide on how realistic is that concept, 
is an old and very interesting problem. In this respect, the conformal 
transformation (30) provides a direct way of linking the laws of motion of an 
idealized point-particle with those determining the hydrodynamic flow of a 
realistic volume element in the interior of a continuous source. In the latter 
case, the problem of {\em backreaction} is directly involved in the structure 
of the corresponding stress-energy tensor. To which extend is this procedure 
applicable to any form of ${\cal T}_{ml}$, remains an open problem.

\vspace{0.8cm}

{\bf Acknowledgements:} The authors wish to thank Professor P. S. Florides of 
Trinity College, Dublin University and Dr D. Kazanas of Goddard Space Flight 
Center, NASA, for many illuminating discussions. The authors also thank the 
Greek Secretariat for Research and Technology, for partially supporting this 
research, under the grand PENED No 797.

\section*{References}

\begin{itemize}

\item[] Anderson J. 1967, {\em Principles of Relativity Physics} (New 
York: Academic Press)
\item[] Antonucci R. 1993, ARA\&A 31, 473
\item[] Antonucci R. \& Miller J. 1985, ApJ 297, 621
\item[] Blandford R. \& Rees M. 1992, in {\sl Testing the AGN Paradigm}, 
ed. S. Holt, S. Neff and M. Urri (New York: AIP), 3
\item[] Chandrasekhar S. 1965, ApJ 142, 1488
\item[] Chandrasekhar S. 1983, {\em The Mathematical Theory of Black Holes} 
(Oxford: University Press)
\item[] Chokshi A. \& Turner E. 1992, MNRAS 259, 421
\item[] Cotsakis S. 1993, Phys. Rev. D 47, 1437
\item[] Fock V. 1959, {\em The Theory of Space, Time and Gravitation} 
(London: Pergamon Ltd)
\item[] Gradshteyn I \& Ryzhik I 1965, {\em Tables of Integrals, Series and 
Products} (New York: Academic Press)
\item[] Greenhill L., Jiang D., Moran M., Reid M., Lo K. \& Claussen M. 1995,
ApJ 440, 619
\item[] Greenhill L. \& Gwinn C. 1997, ApJ Supplement Series (preprint)
\item[] Guilbert P. \& Rees M. 1988, MNRAS 233, 475
\item[] Hawking S. \& Ellis G. F. R. 1973, {\em The Large Scale Structure 
of Spacetime} (Cambridge: University Press)
\item[] Hawking \& Penrose R. 1970, Proc. R. Soc. London, A314, 529
\item[] Ho L. 1998, in {\sl Observational Evidence for Black Holes in the 
Universe}, ed. S. Chakrabarti (Dordrecht: Kluwer)
\item[] Holt S. Neff S. \& Urri M. (eds.) 1992, in {\sl Testing the AGN 
Paradigm}, (New York: AIP)
\item[] Jaffe W., Ford H., Ferrarese L., Van den Bosch F. \& Connell R. 1993, 
Nature 364, 213
\item[] Kohno K., Kawabe T., Tosaki M. \& Ocumara S. 1996, ApJ 461, L29
\item[] Kormendy J. \& Richstone D. 1995, ARA\&A 33, 581
\item[] Krolik J. \& Begelman M. 1988, ApJ 329, 702
\item[] Maloney P., Begelman M. \& Rees M. 1995, ApJ 432, 606
\item[] Maoz E. 1995, ApJ 447, L91
\item[] Misner C., Thorne K. S. \& Wheeler J. A. 1973, {\em Gravitation} 
(San Francisco: Freeman)
\item[] Miyoshi M., Moran M., Hernstein J., Greenhil L., Nakar N., Plamond P. 
\& Inoue M. 1995, Nature 373, 127
\item[] Narlikar J. 1983, {\em Introduction to Cosmology} (Boston: Jones 
and Bartlett)
\item[] Neufeld D. \& Maloney P. 1995, ApJ 447, L17
\item[] Novikov I. \& Thorne K. S. 1973, {\em Astrophysics of Black Holes}, 
in {\sl Black Holes - Les Astres Occlus}, Les Houches, August 1972 (New York: 
Gordon and Breach)
\item[] Papapetrou A. 1974, {\em Lectures on General Relativity} (Dordrecht: 
Reidel Publishing)
\item[] Peebles  P. 1972, ApJ 178, 371
\item[] Penrose R. 1964, {\em Conformal Treatment of Infinity}, in 
{\sl Relativity Groups and Topology}, ed. B. S. De Witt and C. De Witt 
(New York: Gordon and Breach)
\item[] Quinlan G., Hernquist L. \& Sigurdsson S. 1995, ApJ 440, 554  
\item[] Rees M. 1984, ARA\&A 22, 471
\item[] Rees M. 1995, {\em Perspectives in Astrophysical Cosmology} 
(Cambridge: University Press)
\item[] Ryan M. \& Shepley L. 1975, {\em Homogeneous Relativistic Cosmologies} 
(Princeton: University Press)
\item[] Rydbeck G., Wilkind T., Cameron M., Wild W., Eckart A., Genzel R. \& 
Rothermel H. 1993, A\&A 270, L13
\item[] Saglia R. 1996, in Proceedings of the IAU Symposium No 171, {\sl 
New Light on Galaxy Evolution}, ed. R. Bender and R. L. Davies (Dordrecht: 
Kluwer), 157
\item[] Sargent W., Young P., Boksenberg A., Shortridge K. Lynds C. \& 
Harwick F. 1978, ApJ 221, 731
\item[] Schiff I. 1960, Am. J. Physics 28, 340
\item[] Shakura N. \& Sunyaev R. 1973, A\&A 24, 337
\item[] Spyrou N. K. 1976, ApJ 209, 243
\item[] Spyrou N. K. 1997a, in Proceedings of the $18^{th}$ Summer School 
and International Symposium, {\sl The Physics of Ionized Gases}, ed. B. 
Vujicic, S. Djurovic and I. Puric (Institute of Physics, Novi Sad University, 
Jugoslavia), 417
\item[] Spyrou N. K. 1997b, {\sl in Honorem L. N. Mavrides}, eds. G. Asteriadis, 
A. Bandellas, M. Contadakis, K. Katsambalos, A. Papademetriou and I. Tziavos 
(Thessaloniki, Greece) 
\item[] Spyrou N. K. 1997c, {\sl Facta Universitatis} 4, 7
\item[] Spyrou N. K. 1999, in Proceedings of the International Seminar, {\sl 
Current Issues of Astronomical and Planetary Enviromental Concern}, ed. N. K. 
Spyrou (Astronomy Department, Aristoteleion University of Thessaloniki, Greece), 
{\em in press}.
\item[] Spyrou N. K. \& Dionysiou D. 1973, ApJ 183, 265
\item[] Spyrou N. K. \& Varvoglis H. 1982, ApJ 255, 674
\item[] Tacconi L., Genzel R., Blietz M., Cameron M., Harris A. \& Madden S. 
1994, ApJ 426, L77
\item[] Taub A. 1959, Arch. Rat. Mech. Anal. 3, 312
\item[] Urri M. \& Padovani P. 1995, PASP 107, 803
\item[] Weinberg S. 1972, {\em Gravitation and Cosmology} (New York: Wiley \& 
Sons)
\item[] Whitt B. 1984, Phys. Lett. B 145, 176

\end{itemize}

\end{document}